\documentclass[twocolumn]{jpsj2}
\def\ds{\displaystyle}
\def\bm#1{{\mbox{\boldmath $#1$}}}

\def\bms#1{{\mbox{\rm\scriptsize\it\bf#1}}}

\def\a{\alpha}
\def\b{\beta}

\def\e{\varepsilon}

\def\n{\nu}

\def\r{\rho}
\def\s{\sigma}
\def\t{\tau}

\def\w{\omega}

\def\dd{\mbox{d}}

\def\ket{\rangle}

\def\ra{\rightarrow}

\def\ua{\uparrow}
\def\da{\downarrow}

\def\nn{\nonumber}

\title{%
Calculation of Optical Conductivity of YbB$_{12}$ using Realistic Tight-Binding Model
}

\author{%
Tetsuro {\sc Saso}
}

\inst{%
Department of Physics, Faculty of Sciences, Saitama University, Shimo-Ohkubo 255, Sakura-ku, Saitama-City, Saitama 338-8570 Japan.
}

\recdate{June 12, 2004}

\abst{%
Based on the previously reported tight-binding model fitted to the LDA+U band calculation, optical conductivity of the prototypical Kondo insulator YbB$_{12}$ is calculated theoretically.  Many-body effects are taken into account by the self-consistent second order perturbation theory.  The gross shape of the optical conductivity observed in experiments are well described by the present calculation, including their temperature-dependences.
}

\kword{%
YbB$_{12}$, Kondo insulator, tight-binding model, optical conductivity, dynamical conductivity
}

\begin{document}
\maketitle

\section{Introduction}
Many compounds including the rare-earth and actinide ions exhibit the metallic behaviors with heavy effective mass.  Such behaviors can be understood by the combination of the local Kondo-like effect due to the strongly correlated magnetic ions and the periodicity of their alignment.
Some of them, however, become insulator at low temperatures.  Examples can be found in SmB$_6$,\cite{SmB6} YbB$_{12}$,\cite{YbB12} Ce$_3$Bi$_4$Pt$_3$,\cite{Ce3Bi4Pt3} CeRhSb,\cite{CeRhSb} CeFe$_4$P$_{12}$,\cite{CeFe4P12} CeNiSn,\cite{CeNiSn} TmSe,\cite{TmSe} etc.
Although they exhibit insulating behavior at low temperatures, the Kondo-like behaviors (the enhanced electronic specific heat and the paramagnetic susceptibility, etc.) are often observed at higher temperatures, too.  These materials are called Kondo insulators, Kondo semiconductors or heavy fermion semiconductors.\cite{Riseborough00}  

The simplest theoretical model to describe the Kondo insulators is the periodic Anderson model (PAM), which describes the hybridization between the f and the conduction electrons.  However, if one does not take proper account of the orbital degeneracy of f states nor the conduction band degeneracy, the hybridization model can not correctly yield an energy gap.\cite{Ohkawa84,Saso03}

In our previous paper,\cite{Saso03} we proposed a band model which takes account of the band and the f-state degeneracies for the most typical Kondo insulator, YbB$_{12}$.
It is the simple tight-binding model constructed from the (dd$\sigma$) overlapping integral between 5d$\varepsilon$ ($xy$, $yz$ and $zx$) orbitals on Yb ions, and the four-fold 4f states, $\Gamma_8$, under cubic crystal field.
This (dd$\sigma$) should be regarded as the effective hopping through the B$_{12}$ clusters.  
The energy dispersions of the conduction bands are given by the following simple expressions:
\begin{equation}
  E_\bms{k}^{\alpha\beta} = E_{d\varepsilon}+3(\rm{dd}\sigma)\cos(k_\alpha/2)\cos(k_\beta/2),
  \label{eq:dxy-band}
\end{equation}
where $(\a,\b)=(x,y)$, $(y,z)$ and $(z,x)$.
The lowest band along $\Gamma$-K-X(110) and $\Gamma$-X(100) are doubly degenerated (in addition to the spin degeneracy).
Introducing the hybridization between these d bands and the 4f $\Gamma_8$ states, we obtained the energy bands describing the LDA+U calculation rather well.  
It was clarified that the proper account of both the conduction band and the f-state degeneracies are indispensable for the correct explanation of the opening of the energy gap in YbB$_{12}$.

In the present paper, we will apply the same model to the calculation of the optical conductivity of YbB$_{12}$.  A possible form of the Coulomb interactions is discussed, and the effect is taken into account via the self-consistent second-order perturbation theory together with the local approximation.  Some problems inherent in the calculation of the optical conductivity are also discussed.  Then the calculation of the optical conductivity spectra will be shown, which exhibits notable temperature-dependences as a result of the many-dody effect.
Comparison will be made with the experimental results by Okamura, et al.\cite{Okamura}  Future problems are discussed in the last section.

\section{Tight-Binding Energy Bands}
In eq.(\ref{eq:dxy-band}), we locate the energy level of 5d$\varepsilon$ orbitals at $E_{\rm{d}\varepsilon}$=1.0 Ryd, and set (dd$\sigma$)= 0.06 Ryd.
Usually, (dd$\sigma$) is negative, but it is set positive here in order to reproduce the LDA+U band calculation.  This choice is justified since the hopping through B$_{12}$ clusters reverses the sign of the effective overlapping integral.

\begin{figure}[t]
\begin{center}
\includegraphics[width=8cm]{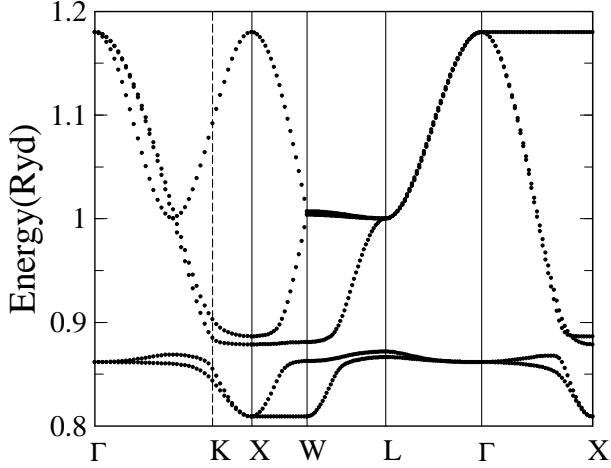}
\end{center}
\caption{The tight-binding band for YbB$_{12}$.  The X points in the left and right denote the equivalent (110) and (100) points (in unit of $2\pi/a$), respectively.}
\label{Fig:Band}
\end{figure}

\begin{figure}[t]
\begin{center}
\includegraphics[width=8cm]{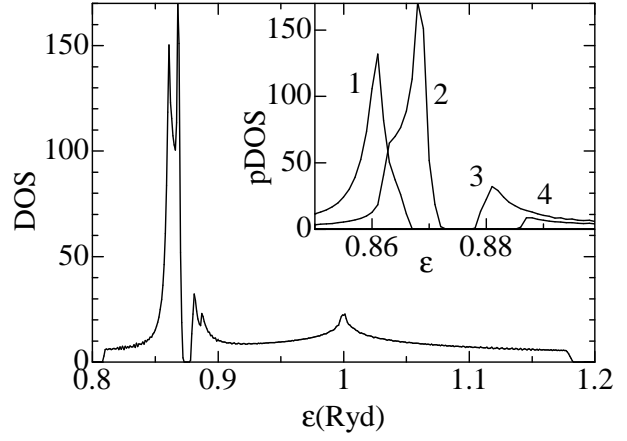}
\end{center}
\caption{The density of states calculated from the tight-binding band for YbB$_{12}$.  The inset shows the contribution from each band (partial DOS) near the gap.  The band index number is attached to each curve.}
\label{Fig:dos0}
\end{figure}

The hybridization between these 5d bands and the 4f states is described by the effective (df$\sigma$) integrals between the nearest-neighbor Yb sites.
These matrix elements (Slater-Koster integrals\cite{Slater54}) have so far been given only for the d and f states under cubic CEF and without the spin-orbit interaction.\cite{Takegahara80}  
For an efficient numerical calculation in later sections, we need the direct expressions for the hybridization between the f-states with the spin-orbit interaction and the d-states under the cubic crystalline field.
Namely, the $\Gamma_8$ states under cubic CEF in the subspace of the total angular momentum $J=7/2$ (for Yb) are expressed in terms of the spherical harmonics $Y_\ell^m$'s with $\ell=3$ and the spinors $\chi_\pm$'s.
Then we write down the expression for $|\Gamma_8^{(1,2)}\pm\ket$ in terms of the $\ell=3$ cubic harmonics states (A$_{2u}$: $|xyz\ket$, T$_{1u}$: $|x(5x^2-3r^2)\ket$, $|y(5y^2-3r^2)\ket$, $|z(5z^2-3r^2)\ket$, T$_{2u}$:  $|x(y^2-z^2)\ket$, $|y(z^2-x^2)\ket$,  $|z(x^2-y^2)\ket$).
Finally, using the mixing matrix elements between the cubic harmonics for d and f states given in ref.\citen{Takegahara80},
 we obtain the necessary matrix elements between the d and f states at $\bm{k}$ point as follows:
\begin{equation}
    \begin{array}{c|cc}
      & \Gamma_8^{(1)}+ & \Gamma_8^{(1)}-  \\
      \hline
      xy\ua & 5it_1(c_xs_y-is_xc_y) & 0  \\
      yz\ua & -4t_1c_ys_z & it_1s_yc_z  \\
      zx\ua & -4it_1s_zc_x & it_1c_zs_x  \\
      xy\da & 0 & -5it_1(c_xs_y+is_xc_y)  \\
      yz\da & it_1s_yc_z & -4t_1c_ys_z  \\
      zx\da & it_1c_zs_x & 4it_1s_zc_x  \\
    \end{array}
\label{eq:H}
\end{equation}
and
\begin{equation}
    \begin{array}{c|cc}
       & \Gamma_8^{(2)}+ & \Gamma_8^{(2)}- \\
      \hline
      xy\ua  & 0 & it_2(c_xs_y+is_xc_y) \\
      yz\ua  & -3it_2s_yc_z & 2t_2c_ys_z \\
      zx\ua  & 3it_2c_zs_x & -2it_2s_zc_x \\
      xy\da  & -it_2(c_xs_y-is_xc_y) & 0 \\
      yz\da  & 2t_2c_ys_z & -3it_2s_yc_z \\
      zx\da  & 2it_2s_zc_x & 3it_2c_zs_x \\
    \end{array}
\label{eq:H}
\end{equation}
where $c_\a=\cos(k_a/2)$, $s_\a=\sin(k_\a/2)$ ($\a=x, y, z$), $t_1=\sqrt{5/56}({\rm df}\sigma$) and $t_2=\sqrt{15/56}({\rm df}\sigma$).  Note that we have retained only the nearest neighbor (df$\sigma$) bonds as the simplest model.  
We locate the $\Gamma_8$ states at $E_{\Gamma_8}=0.88$ Ryd and choose (df$\sigma$)=0.015 Ryd.
We also include (df$\pi)=-0.0075$ Ryd and (ff$\sigma)=-0.003$ Ryd.  These transfer integrals are chosen slightly larger than those in our previous paper\cite{Saso03} to fit the experiments on the optical conductivity.
Furthermore, the filled bands below the gap are shifted down by $\Delta E=-0.011$ Ryd (which is also larger than in ref.\citen{Saso03}) relative to the bands above the gap.  This treatment is in accord with the spirit of the LDA+U treatment.
Using these parameters, we obtain the dispersion curves shown in Fig.\ref{Fig:Band} which have an indirect gap of about 0.0069 Ryd between X and L points and the direct gap of 0.018Ryd. 
These values are larger than the experiments.  The reason will be explained later.
The bands are labeled as 1 to 5 from the bottom to the top, each of which has the Kramers degeneracy.
The density of states (DOS) and the partial DOS are shown in Fig.\ref{Fig:dos0}.

\section{Coulomb Interactions}
We consider the effective tight-binding model mentioned above as a free (mean field) system, and introduce the effective Coulomb and exchange interactions as follows:\cite{Parmenter}
\begin{eqnarray}
\label{eqn:hamiltonian}
{\cal H}  &=& {\cal H}_{\rm band} + {\cal H}_{\rm int} , \\
\label{eqn:H_band}
{\cal H}_{\rm band} &=& 
\sum_{\bms{k}\a\sigma}
E_{\bms{k}\alpha}
c^+_{\bms{k}\alpha\sigma}c_{\bms{k}\alpha  \sigma}, \\
{\cal H}_{\rm int} &=& U \sum_{i\a}
n_{i\alpha \uparrow}n_{i\alpha \downarrow} 
+ U_{2}\sum_{i, \alpha > \alpha',\sigma}
n_{i\alpha \sigma}n_{i\alpha' -\sigma} \nonumber \\
& & + U_{3} \sum_{i, \alpha > \alpha',\sigma}
n_{i\alpha \sigma}n_{i\alpha' \sigma} \nonumber \\
& & -J \sum_{i, \alpha > \alpha',\sigma}
c^+_{i\alpha \sigma}c_{i\alpha -\sigma}
c^+_{i\alpha' -\sigma}c_{i\alpha' \sigma},
\label{eqn:H_int} 
\end{eqnarray}
where
$E_{\mbox{\scriptsize\it\bf k}\a}$ denotes the diagonalized tight-binding band energy 
for the $\a$-th band. 
We denote the annihilation (creation) operator for the  $\a$-th band
in the momentum and the site representations as $c_{\bms{k} \a \sigma}$ ($c^+_{\bms{k} \a \sigma}$) and $c_{i \a \sigma}$ ($c^+_{i \a \sigma}$), respectively. 
Here the index $\sigma$ denotes a pair of the time-reversal states which we call spin hereafter.
The parameters $U$, $U_{2}$, $U_{3}$ and $J$ correspond to the intra-band, the inter-band anti-parallel spin, the inter-band parallel spin Coulomb interactions, and the inter-band exchange interaction,
respectively.  
These parameters should be taken as the effective (reduced) ones because of the uncertainty due to the LDA+U treatment and the renormalization of the higher-order processes of interactions.  We regard their values as the low energy limits of the renormalized interactions.

%%%%%%%%%%%%%
%In the Hamiltonian (\ref{eqn:H_int}), the Coulomb and exchange interactions are introduced bewteen all bands,  although
%the strong electron-electron interactions should exist only between the f electrons in real systems.
%%%%%%%%%%%%%
In the Hamiltonian (\ref{eqn:H_int}), the Coulomb and exchange interactions are introduced bewteen the band-diagonalized states.
We retained only the simplest terms and neglected the more complicated matrix elements.
Furthermore, we neglected the orbital dependence of the interaction parameters and assumed that the interactions act on all the orbitals equally.
Of course, the Coulomb repulsion acts most strongly on the f-states.
%%%%%%%%%%%%%
However, the interactions act more strongly on the electrons in the region of the high density of states, so that the f-componets will be more affected than the others even in this form.
The other reason to choose the above form is that if we use the extended periodic Anderson-like model and introduce the interaction on the f-states only, the Green's function, and hence the density of states, can not be expressed by the DOS of the unperturbed Hamiltonian.
Instead, one has to perform the three-dimensional $\bm{k}$-summation in each time of the iterative calculation as
\begin{equation}
  G_\a^f(\bm{k},\e) = \frac{1}{N}\sum_\bms{k}\frac{1}{\ds \e-E_{f\a}-\Sigma_\a^f(\bm{k},\e)-\sum_\n\frac{|V_{\bms{k}\n\a}|^2}{\e-\e_{\bms{k}\n}}},
\label{eq:GreenFn}
\end{equation}
to obtain the local f-electron Greenian or the renormalized local DOS, unless the the hybridization $V_{\bms{k}\n\a}$ can be expressed merely by $\e_{\bms{k}\n}$, the energy of the conduction electrons before the hybridization is introduced.

On the other hand, the above form of the interactions in (\ref{eqn:H_int}) allows one to express the local Green's function for the $\a$-th band by the unperturbed DOS of the $\a$-th band $\r_\a^0(\e)$:
\begin{eqnarray}
\label{eqn:Green}
G_{\a}(\e) &=&
\int {\rm d}\nu \frac{\r_\a^0(\nu)}
{\e + i \delta-\nu-\Sigma_{\a}(\e)}, \\
\r_\a^0(\nu) &=&  \frac{1}{N}
\sum_{\mbox{\scriptsize\it\bf k}}\delta(\nu-E_{\mbox{\scriptsize\it\bf k}}^{\a}),
\end{eqnarray}
where $\Sigma_{\a}(\e)$ denotes the self-energy for the $\a$-th band (diagonal in $\a$ in the present model) and is assumed $\bm{k}$-independent (the local approximation), $N$ the number of sites and $\delta\rightarrow 0^+$.
Since the three-dimensional $\bm{k}$-summation is needed only once at the beginning for the calculation of $\r_\a^0(\n)$, the present treatment reduces the computational time to the tractable range even for realistic systems.  

For simplicity and in order to reduce the number of parameters,
the following relations are further assumed; $U_{2}=U-J$ and
$U_{3}=U-2J$ \cite{Parmenter}. 
In this case, ${\cal H}_{\rm int}$ can be expressed as
\begin{eqnarray}
{\cal H}_{\rm int} &=&
\frac{U}{2}\sum_{i\a\a'\sigma\sigma'}
c^+_{\a' i \sigma'} c^+_{\a i \sigma}
c_{\a i \sigma} c_{\a' i \sigma'} \nonumber \\
&-& \frac{J}{2}\sum_{i\a\a'\sigma}
c^+_{\a' i \sigma}c^+_{\a i \bar{\sigma}}
c_{\a i \bar{\sigma}}c_{\a' i \sigma},
\end{eqnarray}
which has a rotational symmetry in both the orbital and the spin states.

To take account of the correlation effect, 
we apply the self-consistent second-order perturbation theory (SCSOPT) to the present model together with
the local approximation for the self-energy part of the Green's function. (See ref.\citen{Urasaki99,Saso97} for example.)
Renormalized (reduced) values will be assumed for the interaction parameters, so that we expect that
the SCSOPT may be valid to investigate the correlation effect in the low-energy and low temperature region.

\begin{figure}[t]
\begin{center}
\includegraphics[width=8cm]{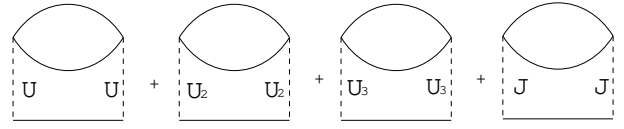}
\end{center}
\caption{The second-order processes for the self-energy are shown.}
\label{Fig:SelfEnergy}
\end{figure}

The self energy $\Sigma_{\a}(\e)$ consists of the constant Hartree term
and the second-order perturbation term $\Sigma_\a^{(2)}(\e)$. 
The former is absorbed into the chemical potential, which is determined at each temperature to compensate the electron and hole numbers.
The latter is calculated from the full local Green's function as shown in Fig.\ref{Fig:SelfEnergy} and given in the imaginary time expression by
\begin{eqnarray}
  \Sigma_\a^{(2)}(\t)&=&U^2G_\a(\t)^2G_\a(-\t) \nn \\
  & & \hspace{-2cm}+(U_2^2+U_3^2+J^2)G_\a(\t)\sum_{\a'\neq\a}G_{\a'}(\t)G_{\a'}(-\t).
\end{eqnarray}
The actual calculation is performed on the real energy axis.\cite{Saso97,Saso01}
The quasi-particle density of states is calculated by $\rho_{\a}(\e) = -(1/\pi){\rm Im}G_{\a}(\e)$.

It is noted that 
the second-order self-energy disappears at $T\ra 0$ in the present model, since all the carriers die out.  Therefore, the gap is not renormalized at $T=0$.\cite{Urasaki99}  This is in contrast to the periodic Anderson-like model, in which the self-energy is finite at $T=0$, so that the gap size is renormalized to a value of the order of the Kondo temperature.\cite{Saso97}
At finite temperatures in the present model, the self-energy becomes finite and has the imaginary-part due to the scattering between the thermally excited carriers, so that the gap in the quasi-particle DOS is filled up gradually.  Effects will be more enhanced in the optical conductivity (see eq.(\ref{eq:opcon-JDOS})).

\section{Optical Conductivity}
\begin{figure}[t]
\begin{center}
\includegraphics[width=8cm]{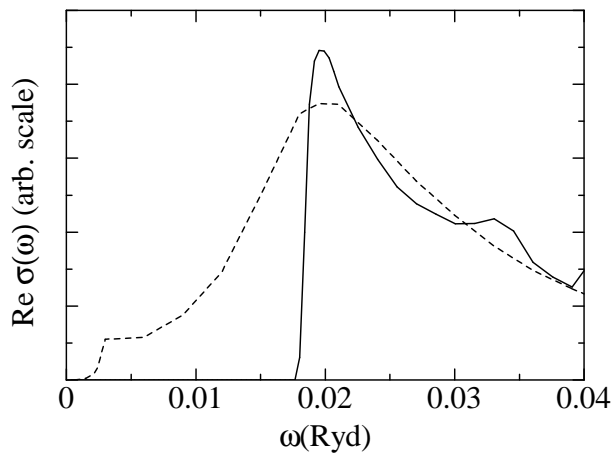}
\end{center}
\caption{The frequency dependence of the optical conductivity for $U=J=0$ and $T=0$ based on the direct-transition formula (\ref{eq:opcon-direct}).  The dotted line indicates the observed spectrum at $T=20$ K.\cite{Okamura}}
\label{Fig:OptU0T0}
\end{figure}

The optical conductivity is usually calculated by the Kubo formula
\begin{eqnarray}
  \mbox{Re} \s(\w) &\propto& \int_{-\infty}^\infty \dd\e \frac{f(\e)-f(\e+\w)}{\w} \nn \\
  & & \hspace{-1cm} \times \sum_{\bms{k}\a\b}|\bm{v}_\bms{k}^{\a\b}|^2\mbox{Im}G_\bms{k}^\a(\e)\mbox{Im}G_\bms{k}^\b(\e+\w),
\label{eq:opcon-direct}
\end{eqnarray}
where $\bm{v}_\bms{k}^{\a\b}$ is the velocity matrix element and $\a$, $\b$ denote the band indices.  Note that the momentum $\bm{k}$ is conserved in this formula since only the direct transitions are included here. 
%%%%%%%%%%%
It is well known that this formula can describe the optical conductivity of ordinary semiconductors well.\cite{Chelikowsky76}
%%%%%%%%%%%
Neglecting the $\bm{k}$-dependence of $\bm{v}_\bms{k}^{\a\b}$, we have calculated this formula for the tight-binding band constructed in \S 2.  
The result is displayed in Fig.\ref{Fig:OptU0T0}, which does not reproduce the experiment\cite{Okamura} (the dotted line in the same figure) at all.

%%%%%%%%%%%
%Mahan wrote in his famous textbook\cite{Mahan00} that the formula (\ref{eq:opcon-direct}) is a very definite prediction, but is not observed (in experiments).
%In the real situations, indirect transitions may occur due to impurities, phonon-assisted process, etc., and the threshold shape of the absorption is shifted and enhanced due to the effects of the exciton absorption and the local Coulomb potential (Sommerfeld factor\cite{Haug} $C(\w)\propto (\hbar\w-E_g)^{-1/2}$  at the threshold).   
%%%%%%%%%%%
Mahan wrote in his famous textbook\cite{Mahan00} that the formula (\ref{eq:opcon-direct}) is a very definite prediction, but is not observed (in experiments), warning an importance of the effect of the Coulomb interaction (excitons).
In addition, indirect transitions may occur due to impurities, phonon-assisted process, etc. in the real situations.
Especially, the impurity effects may not be neglected in YbB$_{12}$ since the resistivity increases at low temperatures but saturates at finite values.\cite{Sugiyama88}  The threshold shape of the absorption is modified by the exciton absorption and enhanced by the Sommerfeld factor\cite{Haug} $C(\w)\propto (\hbar\w-E_g)^{-1/2}$  at the threshold).   
%%%%%%%%%%%
It was also pointed out that the mutual Coulomb interaction yields an effect similar to the indirect transitions.\cite{Logan03}  

\begin{figure}[t]
\begin{center}
\includegraphics[width=8cm]{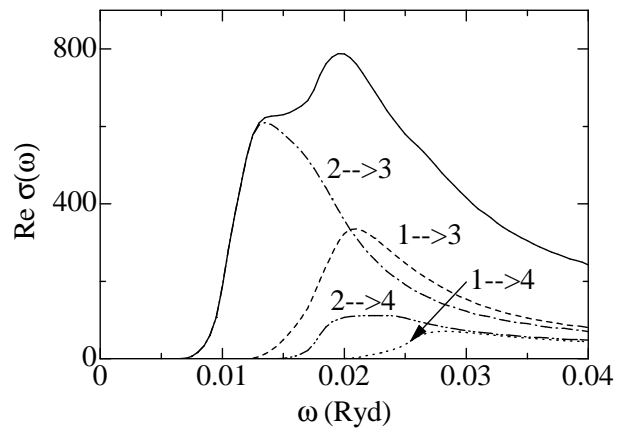}
\end{center}
\caption{Contribution of each band-to-band transition to the optical conductivity when $T=U=J=0$.}
\label{Fig:OptJU0}
\end{figure}

We have previously experienced\cite{Urasaki99} that the neglect of the momentum conservation yields a much better agreement with experiments in the case of FeSi (another Kondo insulator with the transition metal element instead of the rare-earth).  
Therefore, we use the joint-DOS-type formula
\begin{equation}
  \mbox{Re}\s(\w) \propto \sum_{\a\b} \int_{-\infty}^\infty \dd\e \frac{f(\e)-f(\e+\w)}{\w}\r_\a(\e)\r_\b(\e+\w),
\label{eq:opcon-JDOS}
\end{equation}
which completely neglects the momentum conservation.
%%%%%%%%%%%%%%%
Use of this formula is partially motivated by the possible violation of the momentum conservation mentioned above.
%%%%%%%%%%%%%%%
This formula is not correct in the $\w\ra 0$ limit of metals, but yielded a reasonable result for finite frequencies.\cite{Urasaki99}
We show the contribution of each transition between the bands for $U=J=0$ and $T=0$ in Fig.\ref{Fig:OptJU0}.
Compared with the experiment at the lowest temperature $T=20$K (Fig.\ref{Fig:OptU0T0}, dotted line), the contribution from the $2\ra 3$ band transition is slightly lower in position, making a redundant shoulder at 0.013 Ryd, but the gross structure including the position of the mid-IR peak is reproduced.  
We tried various combination of the tight-binding parameters, and found impossible to simultaneously reproduce (1) the tail extending to the low frequency part $\w <$ 0.01 Ryd (or the threshold of the gap at 0.0022 Ryd), (2) the mid-IR peak width $\sim$ 0.02 Ryd, and (3) the mid-IR peak position at 0.02 Ryd.
Namely, if we fit the calculation to the observed threshold of the spectrum at 0.0022Ryd and the position of the mid-IR pea at 0.02 Ryd, we have to choose a much larger value of (ff$\s$) and hence a larger width of the mid-IR peak.
In other words, a long low-frequency tail can not be reproduced within the present tight-binding model.  Another possible reasons for this disagreement might be the present scheme for the treatment of the many-body interaction, the present formula (\ref{eq:opcon-JDOS}) for the optical conductivity, or the neglect of the impurity effect.
Despite these deficiencies, we consider that the present choice of the parameters are the best within the present model and the scheme.

\section{Comparison with Experiments}
\begin{figure}[t]
\begin{center}
\includegraphics[width=8cm]{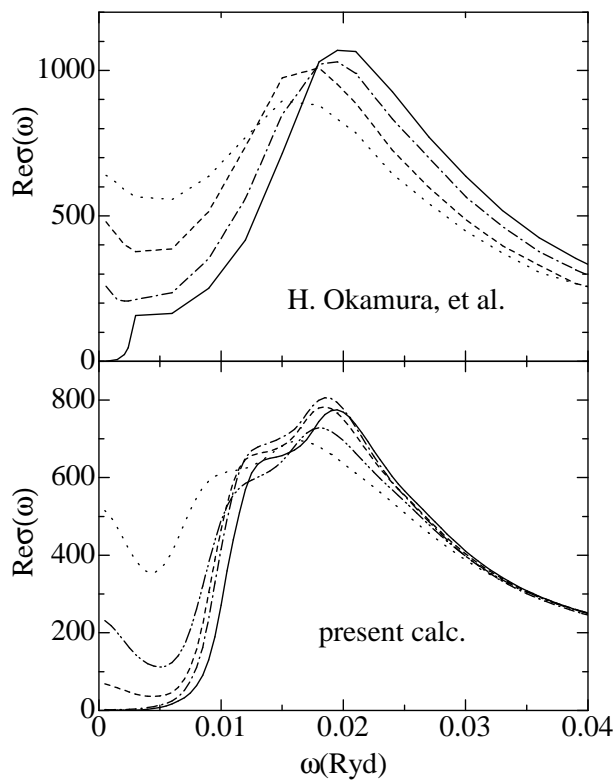}
\end{center}
\caption{Lower panel: the calculated temperature dependence of the optical conductivity spectra for $U=0.02$ Ryd and $T=$0, 80, 160, 220 and 290 K, respectively).  Upper panel: Corresponding experimental results for $T=$20, 78, 160 and 290 K by H. Okamura, et al.\cite{Okamura}}
\label{Fig:OptU02T}
\end{figure}

Optical conductivity spectra measured by Okamura, et al.\cite{Okamura} at finite temperatures, $T=$20, 78, 160 and 290 K, are displayed in the upper panel of Fig.\ref{Fig:OptU02T}, and our theoretical results based on the formula (\ref{eq:opcon-JDOS}) for the corresponding temperatures in the lower panel.   A calculation for an intermediate temperature $T=$220 K is added.
Interaction parameters are chosen as $U=$0.002 Ryd and $J=0.2U$, respectively to fit the experiments.
As was mentioned in the last section, the low frequency parts at $\w<$ 0.01 Ryd are not well fitted, but the over-all spectra and the temperature-variations are reproduced with the present parameters.  Namely, (1) the mid-IR peak originates from the transition between the hybridized f-d bands (often called as the f-d transition in the local picture), (2) the threshold may better correspond to the indirect gap than the direct gap, although an origin of the low frequency part $\w<$ 0.01 Ryd is not clear.

\begin{figure}[t]
\begin{center}
\includegraphics[width=8cm]{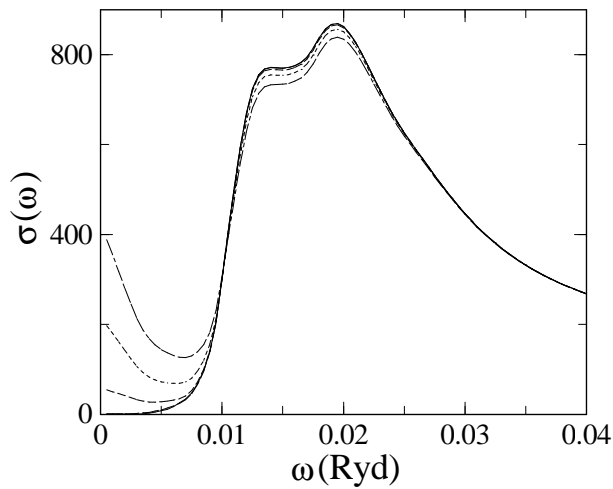}
\end{center}
\caption{The temperature dependence of the optical conductivity for $U=J=0$ and for the same temperature as in Fig.\ref{Fig:OptU02T}.}
\label{Fig:OptU0T}
\end{figure}

Furthermore, the effect of many-body interaction should be underlined here.
Namely, the gap $E_g\sim$ 350K seen in the observed spectra at the lowest temperature is filled up quickly at $T=78$ K and the mid-IR peak is shifted to the lower frequencies.  The present calculation reproduces these features partly, although the gap is filled up more slowly with increasing temperature.
It may be due to the fact that the interaction is not effective at low temperatures in our model Hamiltonian (\ref{eqn:hamiltonian}).
This is also seen in the temperature dependence of the DOS.
%%%%%%%%%%%%%%
Higher order process of interactions and consideration of vertex corrections may be important.
%%%%%%%%%%%%%%
Other important effect of interaction is the shift of the mid-IR peak positions.
They shift to low frequencies as the temperature increases.

To see these effects more in detail, we display the calculation for Re$\s(\w)$ without the self-energy in Fig.\ref{Fig:OptU0T}.  It has a similar temperature dependences with Fig.\ref{Fig:OptU02T} for $T\le$160K but the filling-up of the gap region is slow for higher temperatures.  Namely, the many-body effect becomes effective and noticeable at $T>$160 K in the present calculation.
It should also be noted that the shift of the mid-IR peak can not be explained at all without the interactions as shown in Fig.\ref{Fig:OptU0T}.

\section{Conclusions}
In the present paper, we have calculated the optical conductivity spectra for the most typical Kondo insulator YbB$_{12}$, based on the previously proposed tight-binding model fitted to the LDA+U calculation.
The Coulomb and exchange interaction parameters are introduced onto the states in which the hybridizations are diagonalized.
As a result, the second-order perturbation theory was easily perfomred iteratively for the realistic band model.
The calculated optical conductivity spectra reproduced the gross features of the experiments: 1) The mid-IR peak at 0.02 Ryd, 2)The shift of its position to low frequencies as the temperature increases,  and 3) the gap is filled up faster than expected from the gap size as $T$ increases.

The following issues are to be improved in the future: 
1) The low energy part $\w<$ 0.01 Ryd of the spectra is not reproduced.
2) The temperature variation of the spectra is weaker than the experiments.  
The cause of the latter will be that the interactions are not effective enough at low temperatures in the present model Hamiltonian.

In the present paper, we have neglected the effect of the matrix element $\bm{v}_\bms{k}^{\a\b}$ in the optical conductivity.
It is important to take this factor into account for more realistic calculation, but we point out that the local symmetries are mixed up at the general $\bm{k}$-point.  Thus the neglect of the matrix element may not be crucial.

Despite these issues to be solved, the present direction of the reasearch will be very useful for the  study of the strongly correlated systems.  Recently, proposal has been made on the unified ab initio treatment of the LDA band calculation and the correlation effect.\cite{Kotliar01} 
We consider, however, a more practical approach starting from a simple but realistic tight-binding model followed by the self-energy calculations will be still useful in the present stage.
It is important to construct a theory with correlation effect, which can consistently explain all the thermal,\cite{YbB12} thermoelectric,\cite{Iga} transport\cite{Sugiyama88} and magnetic\cite{Alekseev02} properties of YbB$_{12}$ based on the curren approach.

To improve the present calculation further, one has to include the full anisotropy in the $\bm{k}$-space and perform the three-dimensional $\bm{k}$-summation in each iteration steps based on eq.(\ref{eq:GreenFn}) and an approapriate self-energy.   
%%%%%%%%%%
%The imaginary part of $\Sigma_\a^f(\bm{k},\e)$ produces the indirect transitions, so that the optical absorption starts at the indirect transition threshold.  
%Such a calculation, if possible, may improve an agreement with the experiments.
%However, such a full calculation is still hard at present.
%On the other hand, a detailed band structure parameters might be determined through the comparison between the theory and experiments, e.g. a precise transport measurement at high magnetic field.\cite{Iga99}
%%%%%%%%%%
The correlation effects may be taken into account beyond the SCSOPT via the FLEX approximation.\cite{Bickers89}  A vertex correction to eq.(12) and (13) will be also important.
Such higher processes may produce the indirect transitions (possiblly combined with disorder), rendering the optical absorption at the indirect transition threshold.  
These calculations, if possible, may improve an agreement with the experiments,
but full calculations are still hard at present.
On the other hand, the present band model must be also improved.  Precise band structure parameters might be determined through the comparison between the theory and experiments, e.g. a detailed transport measurement at high magnetic field.\cite{Iga99}

%
% for acknowledgement
%
\acknowledgement
The author thanks Professors H. Okamura, T. Mutou and M. Sakai for useful communication and conversation.  This work is supported by the Grant-in-Aid for Scientific Research on
Basic Research (B)(1), ``Development of General Band Calculation Program by the Dynamical Mean-Field Thory'', No. 14340108 and 
Priority Areas, ``Evolution of New Quantum Phenomena Realized in the Filled Skutterudite Structure'', No. 16037204
from the Ministry of Education, Science and Culture.

\end{document}